\documentclass[a4paper,11pt]{article}
\usepackage{pos}
\usepackage{subfigure}

\newcommand{\be}{\begin{eqnarray}}
\newcommand{\ee}{\end{eqnarray}}
\newcommand{\nee}{\nonumber\end{eqnarray}}

\newcommand{\drbar}{{\overline{\rm DR}}}

\newcommand{\mch}[1] {m_{\ti \x^+_{#1}}}
\newcommand{\mnt}[1] {m_{\ti \x^0_{#1}}}

\newcommand{\msg}    {m_{\ti g}}
\newcommand{\msu}[1] {m_{\ti u_{#1}}}
\newcommand{\msd}[1] {m_{\ti d_{#1}}}

\def\be            {\begin{equation}}
\def\ee            {\end{equation}}
\def\bea            {\begin{eqnarray}}
\def\eea            {\end{eqnarray}}

\def\l               {\lambda}

\def\x               {\chi}
\def\ti              {\tilde}

\def\st              {\ti t}
\def\sc              {\ti c}
\def\sbot            {\ti b}
\def\ch              {\ti \x^\pm}

\def\nt              {\ti \x^0}
\def\sg              {\ti g}

\def\sneut           {\ti \nu}

\def\su                  {\ti{u}}

\def\sd                  {\ti{d}}
\def\ss                  {\ti{s}}

\definecolor{darkgreen}{rgb}{0,.5,0}

\title{\bf Imprint of quark flavor violating SUSY 
in h(125) decays at future lepton colliders
}
\ShortTitle{Imprint of QFV SUSY in h(125) decays}

\author*[a]{Keisho Hidaka}
\author[b]{Helmut Eberl}
\author[b]{Elena Ginina}

\affiliation[a]{Department of Physics, Tokyo Gakugei University,\\
  Koganei, Tokyo 184-8501, Japan}

\affiliation[b]{Institut f\"ur Hochenergiephysik der \"Osterreichischen Akademie
  der Wissenschaften, \\ 
  A-1050 Vienna, Austria}
  
\emailAdd{hidaka@u-gakugei.ac.jp}
\emailAdd{helmut.eberl@oeaw.ac.at}
\emailAdd{elena.ginina@oeaw.ac.at}

\abstract{
We study the CP-even neutral Higgs boson decays $h \to c \bar{c}, b \bar{b}, 
b \bar{s}, \gamma \gamma, g g$ in the Minimal Supersymmetric Standard Model (MSSM) 
with general quark flavor violation (QFV), identifying the h as the Higgs boson 
with a mass of 125 GeV. 
We compute the widths of the h decays to $c \bar c, b \bar b, b \bar s (s \bar b)$ 
at full one-loop level. For the loop-induced h decays to photon photon 
and gluon gluon we compute the widths at NLO QCD level. 
{\it For the first time}, we perform a systematic MSSM parameter scan 
including Supersymmetric (SUSY) QFV parameters respecting 
all the relevant constraints, i.e. theoretical constraints from vacuum stability 
conditions and experimental constraints, such as those from K- and B-meson data,  
electroweak precision data, and the 125 GeV Higgs boson data from recent LHC 
experiments, as well as the limits on SUSY particle 
masses from the LHC experiment. We also take into account the expected SUSY 
particle mass limits from the future HL-LHC experiment in our analysis. 
{\it In strong contrast to} the usual studies in the MSSM with quark flavor 
conservation, we find that the deviations of these MSSM decay widths from 
the Standard Model (SM) values can be quite sizable and that there are 
significant correlations among these deviations. 
All of these sizable deviations in the h decays are due to (i) large 
scharm-stop mixing and large scharm/stop involved trilinear couplings 
$T_{U23}, T_{U32}, T_{U33}$, (ii) large sstrange-sbottom mixing and large 
sstrange/sbottom involved trilinear couplings $T_{D23}, T_{D32}, T_{D33}$ 
and (iii) large bottom Yukawa coupling $Y_b$ for large $\tan\beta$ and 
large top Yukawa coupling $Y_t$. 
Future lepton colliders such as ILC, CLIC, CEPC, FCC-ee and MuC can 
observe such sizable deviations from the SM at high signal significance 
{\it even after} the failure of SUSY particle discovery at the HL-LHC.
In case the deviation pattern shown here is really observed at the lepton 
colliders, then it would strongly suggest the discovery of QFV SUSY (the 
MSSM with general QFV). 
}

\FullConference{%
  41st International Conference on High Energy physics - ICHEP2022\\
  6-13 July, 2022\\
  Bologna, Italy
}

\begin{document}

\begin{flushright}
 HEPHY-PUB 1024/22\\
\end{flushright}

\maketitle

\section{Introduction}
What is the SM-like Higgs boson discovered at LHC? 
It can be the SM Higgs boson. 
It can be a Higgs boson of New Physics. 
This is one of the most important issues in the present particle physics field. 
Here we study a possibility that it is the lightest Higgs boson $h^0$ of the 
Minimal Supersymmetric Standard Model (MSSM), focusing on the decays $h^0(125) 
\to c \bar c, b \bar b , b \bar s, \gamma \gamma, g g$. This work is based on the 
update of our previous papers \cite{h02cc,h02bb,h02gagagg} and contains 
substantial new findings. 

\section{Key parameters of the MSSM}
%
Key parameters in this study are Supersymmetric (SUSY) quark flavor 
violating (QFV) parameters 
$M^2_{Q_{u}23} (\simeq M^2_{Q23})$, $M^2_{U23}$, $T_{U23}$, $T_{U32}$, 
$M^2_{Q23}$, $M^2_{D23}$, $T_{D23}$ and $T_{D32}$ which describe 
the $\ti{c}_L - \ti{t}_L$, $\ti{c}_R - \ti{t}_R$, $\ti{c}_R - \ti{t}_L$, 
$\ti{c}_L - \ti{t}_R$, $\ti{s}_L - \ti{b}_L$, $\ti{s}_R - \ti{b}_R$, 
$\ti{s}_R - \ti{b}_L$, and $\ti{s}_L - \ti{b}_R$ mixing, respectively. 
The quark flavor conserving (QFC) parameters $T_{U33}$ and $T_{D33}$ 
which induce the $\ti{t}_L - \ti{t}_R$ and $\ti{b}_L - \ti{b}_R$ mixing, 
respectively, also play an important role in this study. 
All the parameters in this study are assumed to be real, except the 
CKM matrix. We also assume that R-parity is conserved and that the 
lightest neutralino $\nt_1$ is the lightest SUSY particle (LSP). 

\section{Constraints on the MSSM}
\label{Constraints}
In our study, {\it for the first time}, we perform a systematic 
MSSM-parameter scan including SUSY QFV parameters respecting all 
the relevant constraints, i.e. the theoretical 
constraints from vacuum stability conditions and the experimental 
constraints, such as those from $K$- and $B$-meson data and electroweak 
precision data, as well as the limits on SUSY particle (sparticle) 
masses and the $H^0$ mass and coupling data from LHC experiments. 
Here $H^0$ is the discovered SM-like Higgs boson which we identify 
as the lightest $CP$ even neutral Higgs boson $h^0$ in the MSSM.
The details of these constraints are summarized in Ref. \cite{C7_paper_PRD} 
\footnote{
The recent W boson mass ($m_W$) data from CDF II \cite{CDF} is quite 
inconsistent with the other experimental data. 
This issue of the $m_W$ anomaly is not yet settled. Hence, we do not take 
into account this $m_W$ constraint on the MSSM parameters in our analysis.
}.

\section{Parameter scan}
\label{ParameterScan}
%
We compute the decay widths $\Gamma(h^0 \to c \bar c)$, 
$\Gamma(h^0 \to b \bar b)$ and $\Gamma(h^0 \to b \bar s / \bar b s)$ 
at full 1-loop level and the loop-induced decay widths 
$\Gamma(h^0 \to \gamma \gamma)$ and $\Gamma(h^0 \to g g)$ 
at NLO QCD level in the MSSM with QFV \cite{h02cc,h02bb,h02gagagg}. 
We perform the MSSM parameter scan for these decay widths respecting 
all the relevant constraints mentioned above. 
We generate the input parameter points by using random numbers 
in the ranges shown in Table 1 of Ref. \cite{C7_paper_PRD}. 
All input parameters are $\drbar$ parameters defined at scale 
Q = 1 TeV, except $m_A$(pole) which is the pole mass of the 
$CP$ odd Higgs boson $A^0$. We don't assume a GUT relation for the 
gaugino masses $M_1$, $M_2$, $M_3$.

\indent
From 377180 input points generated in the scan, 3208 points survived 
all the constraints. We show these survival points in the scatter 
plot in this article.

\section{$h^0(125) \to c \bar c, b \bar b , b \bar s$ in the MSSM}
\begin{figure*}[t!]
\centering
  \subfigure[]{
  {\mbox{\resizebox{3.8cm}{!}{\includegraphics{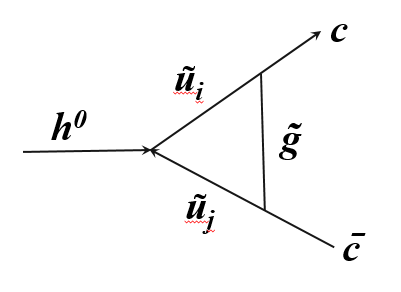}}}}
  \label{h02cc_gluino_loop}}
  \subfigure[]{
  {\mbox{\resizebox{4cm}{!}{\includegraphics{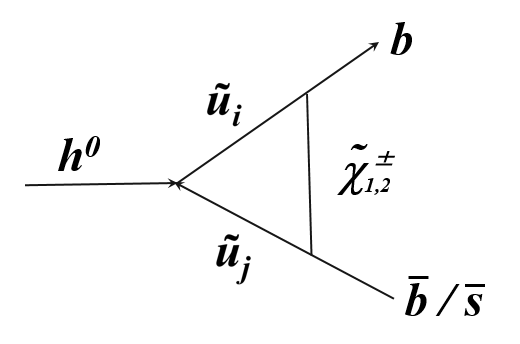}}}}
  \label{h02bb_chargino_loop}}
  \subfigure[]{
  {\mbox{\resizebox{4cm}{!}{\includegraphics{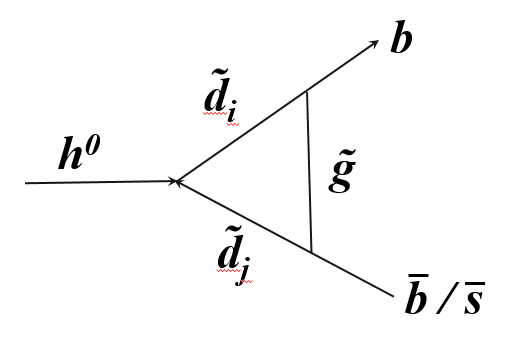}}}}
  \label{h02bb_gluino_loop}}
\caption{
(a) The $\su_i$-$\sg$ loop corrections to $\Gamma(h^0 \to c \bar c)$, 
(b) the $\su_i$-$\ch_{1,2}$ loop and (c) the $\sd_i$-$\sg$ loop corrections 
to $\Gamma(h^0 \to b \, \, \bar b / \bar s)$.
}
\label{1-loop_diag_to_h0_decay}
\end{figure*}

%
%
We compute the decay widths $\Gamma(h^0 \to c \bar c)$, 
$\Gamma(h^0 \to b \bar b)$ and $\Gamma(h^0 \to b \bar s / \bar b s)$ 
at full 1-loop level in the $\overline{DR}$ renormalization scheme 
in the MSSM with QFV using Fortran codes developed by us \cite{h02cc,h02bb}. 
We find that large squark trilinear couplings $T_{U23,32,33}$, 
$T_{D23,32,33}$, large $M^2_{Q23}$, $M^2_{U23}$, $M^2_{D23}$, 
large bottom Yukawa coupling $Y_b$ for large $\tan\beta$, and 
large top Yukawa coupling $Y_t$ can lead to large MSSM 1-loop 
corrections to these widths, resulting in large deviation of 
these MSSM widths from their SM values. \\
%
%
\indent Main MSSM 1-loop corrections to $\Gamma(h^0 \to c \bar c)$ stem 
from the lighter up-type squarks ($\su_{1,2,3}$) - gluino ($\sg$) loops 
at the decay vertex, where $\su_{1,2,3}$ are strong $\sc_{L,R}$ - 
$\st_{L,R}$ mixtures (see Fig. \ref{h02cc_gluino_loop}).  
The large $T_{U23,32,33}$ can enhance the 
$h^0-\su_i-\su_j$ couplings, resulting in enhancement of 
the $\su_i$-$\sg$ loop corrections to $\Gamma(h^0 \to c \bar c)$. \\
%
\indent Main MSSM 1-loop corrections to $\Gamma(h^0 \to b \bar b)$ and 
$\Gamma(h^0 \to b \bar{s}/\bar{b} s)$ stem from 
(i) $\su_{1,2,3}$ - chargino ($\ch_{1,2}$) loops at the decay vertex 
which have $h^0-\su_i-\su_j$ couplings to be enhanced by  
large $T_{U23,32,33}$ (see Fig. \ref{h02bb_chargino_loop}) and 
(ii) $\sd_{1,2,3}$ - $\sg$ loops at the decay vertex, where $\sd_{1,2,3}$ 
are strong $\ss_{L,R}$ - $\sbot_{L,R}$ mixtures (see Fig. \ref{h02bb_gluino_loop}). 
%
%
The large $T_{U23,32,33}$ and $T_{D23,32,33}$ can enhance the 
$\su_i$ - $\ch_{1,2}$ and $\sd_i$ - $\sg$ loop corrections 
to $\Gamma(h^0 \to b \bar b)$, $\Gamma(h^0 \to b \bar{s}/\bar{b} s)$, 
respectively.  

We define the deviation of the MSSM width from the SM width as 
$DEV(X) \equiv \Gamma(h^0 \to X \bar X)_{MSSM}/\Gamma(h^0 \to X \bar X)_{SM} - 1$, (X=c,b).
%
%
DEV(X) is related with the coupling modifier 
$\kappa_X \equiv C(h^0 X \bar X)_{MSSM}/C(h^0 X \bar X)_{SM}$ as $DEV(X)=\kappa_X^2 -1$. 

\begin{figure*}[t!]
\centering
 {\mbox{\resizebox{6.0cm}{!}{\includegraphics{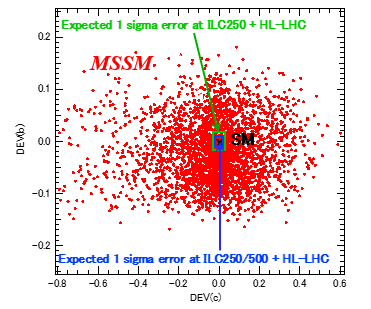}}}}
\caption{
The scatter plot in the DEV(c)-DEV(b) plane obtained 
from the MSSM parameter scan described in Section \ref{ParameterScan}. 
"X" marks the SM point. 
The green and blue box indicate the expected 1$\sigma$ error at 
[ILC250 + HL-LHC] and [ILC250/500 + HL-LHC], respectively \cite{ILC_Higgs}. 
}
\label{DEVc2DEVb}
\end{figure*}

In Fig. \ref{DEVc2DEVb} we show the scatter plot in the DEV(c)-DEV(b) plane 
obtained from the MSSM parameter scan described above, 
respecting all the relevant constraints shown in Section \ref{Constraints}.
We see that DEV(c) and DEV(b) can be quite 
large simultaneously: DEV(c) can be as large as $\sim\pm 60 \%$ and 
DEV(b) can be as large as $\sim\pm 20 \%$.
ILC together with HL-LHC can observe such large deviations from SM at 
high significance \cite{ILC_Higgs}.\\
%
%
\indent We have found that the deviation of the width ratio 
$\Gamma(h^0 \to b \bar b)/\Gamma(h^0 \to c \bar c)$ 
in the MSSM from the SM value can exceed +100\%.\\
\indent From our MSSM parameter scan we find that QFV decay branching 
ratio $B(h^0 \to b s) \equiv B(h^0 \to b \bar{s}) + B(h^0 \to \bar{b} s)$ 
can be as large as $\sim 0.2\%$ (see also \cite{Heinemeyer}) while it 
is almost zero in the SM. The ILC250/500/1000 sensitivity to this 
branching ratio could be $\sim 0.1\%$ at 4$\sigma$ signal significance 
\cite{Tian}. Note that LHC and HL-LHC sensitivity should not be so 
good due to huge QCD background.
%
%
From the scan we find that $B(h^0 \to b s)$ can be large for large 
$|T_{D23}|$ and $|T_{D32}|$ being the size of the $\ti{s}_R - \ti{b}_L$ 
and $\ti{s}_L - \ti{b}_R$ mixing parameter, respectively.

%

%
%

\section{$h^0(125) \to \gamma \gamma, g g$ in the MSSM}
%
We compute the widths of the loop-induced $h^0$ decay to $\gamma \gamma$ 
and $g g$ at NLO QCD level \cite{h02gagagg}. From our MSSM parameter scan, 
we find 
(i) DEV($\gamma$) and DEV($g$) can be sizable simultaneously: 
    DEV($\gamma$) and DEV($g$) can be as large as $\sim \pm1\%$ 
    and $\sim \pm4\%$, respectively. 
(ii) There is a very strong correlation between DEV($\gamma$) 
    and DEV($g$). 
(iii) The deviation of the width ratio 
    $\Gamma(h^0 \to \gamma \gamma)/\Gamma(h^0 \to g g)$ 
    in the MSSM from the SM value can be as large as $\sim \pm5\%$.
(iv) ILC250/500 together with HL-LHC can observe such large 
    deviations from SM at fairly high significance \cite{ILC_Higgs}, 
    except DEV($\gamma$).\\

\section{Benchmark scenario}
\begin{figure*}[t!]
\centering
 {\mbox{\resizebox{6.0cm}{!}{\includegraphics{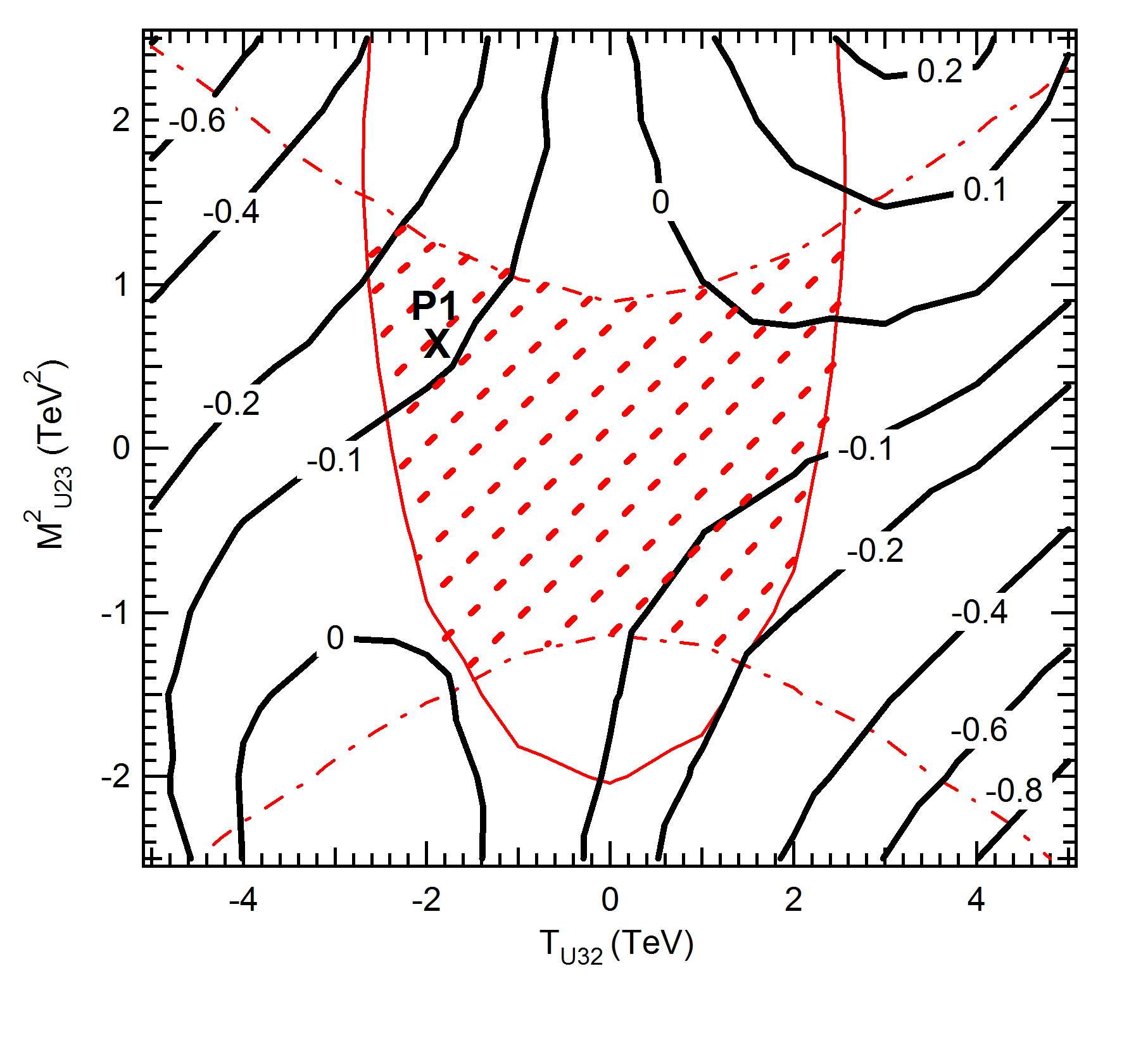}}}}
\caption{
Contour plot of DEV(c) around P1 in the $T_{U32}$-$M^2_{U23}$ plane. 
"X" marks P1. The red hatched region is allowed by all the constraints 
including the expected sparticle mass limits from HL-LHC. 
}
\label{Cnt_DEVc}
\end{figure*}

%
In our analysis we also take into account the expected sparticle mass 
limits from the future HL-LHC experiment. From the allowed MSSM parameter 
points in the scan, we have selected a benchmark point P1 shown in 
Table \ref{Bench_P1} which satisfies also all the expected sparticle 
mass limits (including ($m_{A/H^+}$, $\tan\beta$) limits) from negative 
sparticle-search in the HL-LHC experiment \cite{ESUpgrade, SnowmassRep}. 
The resulting physical masses of the particles are shown in Table \ref{physmasses}. 
In Fig. \ref{Cnt_DEVc} we show the contour plot of DEV(c) around P1 in the 
$T_{U32}$-$M^2_{U23}$ plane. We find that DEV(c) can be very large 
(about -30\% to 10\%) in the sizable region allowed by all the constraints 
including the expected sparticle mass limits from the future HL-LHC 
experiment. For DEV(b), DEV(g) and DEV($\gamma$) we have obtained similar 
results to those presented in Sections 5 and 6. We have also found that 
$B(h^0 \to b s)$ is sizable ($\sim$ 0.1\%) in the allowed region of the 
$T_{U32}$-$M^2_{U23}$ plane. 

%
%
%
%

\begin{table}[h!]
\footnotesize{
\caption{
The MSSM parameters for the reference point P1 (in units 
of GeV or GeV$^2$ except for $\tan\beta$). All parameters are 
defined at scale Q = 1 TeV, except $m_A(pole)$. 
The parameters that are not shown here are taken to be zero.
}
\begin{center}
\begin{tabular}{|c|c|c|c|c|c|}
    \hline
\vspace*{-0.3cm}
& & & & &\\
\vspace*{-0.3cm}
     $\tan\beta$ & $M_1$ &  $M_2$ & $M_3$ &  $\mu$ &  $m_A(pole)$\\ 
& & & & &\\
    \hline
\vspace*{-0.3cm}
& & & & &\\
\vspace*{-0.3cm}
    33 & 1660 & 765 & 4615 & 870 & 5325\\
& & & & &\\
    \hline
    \hline
\vspace*{-0.3cm}
& & & & &\\
\vspace*{-0.3cm}
      $M^2_{Q 22}$ & $ M^2_{Q 33}$ &  $M^2_{Q 23}$ & $ M^2_{U 22} $ & $ M^2_{U 33} $ &  $M^2_{U 23} $\\ 
& & & & &\\
     \hline
\vspace*{-0.3cm}
& & & & &\\
\vspace*{-0.3cm}
     3975$^2$ & 3160$^2$ & 920$^2$ & 3465$^2$ & 1300$^2$ & 795$^2$\\
& & & & &\\
    \hline
    \hline
\vspace*{-0.3cm}    
& & & & &\\
\vspace*{-0.3cm}      
      $ M^2_{D 22} $ & $ M^2_{D 33}$ &  $ M^2_{D 23}$ & $T_{U 23}  $ & $T_{U 32}  $ &  $T_{U 33}$\\ 
& & & & &\\
    \hline
\vspace*{-0.3cm}      
& & & & &\\
\vspace*{-0.3cm}  
      2620$^2$ & 2425$^2$ & -1625$^2$ & -2040 & -1880 & -4945\\
& & & & &\\
 \hline 
\multicolumn{6}{c}{}\\[-3.6mm]  
\cline{1-4}
\vspace*{-0.3cm}      
     & & & \\
\vspace*{-0.3cm}      
     $ T_{D 23} $ & $T_{D 32}  $ &  $ T_{D 33}$ &$T_{E 33} $\\ 
     & & & \\
    \cline{1-4}
\vspace*{-0.3cm}      
     & & & \\
\vspace*{-0.3cm}      
     -2360 & 1670 & -2395 & -300\\
     & & & \\
    \cline{1-4}
\end{tabular}\\[3mm]
\begin{tabular}{|c|c|c|c|c|c|c|c|c|}
    \hline
\vspace*{-0.3cm}      
    & & & & & & & &\\
\vspace*{-0.3cm}      
    $M^2_{Q 11}$ & $M^2_{U 11}$ &  $M^2_{D 11}$ & $M^2_{L 11}$ & $M^2_{L 22}$ & $M^2_{L 33}$ & $M^2_{E 11}$ & $M^2_{E 22}$ & $M^2_{E 33} $\\ 
    & & & & & & & &\\
    \hline
\vspace*{-0.3cm}      
    & & & & & & & &\\
\vspace*{-0.3cm}      
    $4500^2$ & $4500^2$ & $4500^2$  & $1500^2$ & $1500^2$ & $1500^2$ & $1500^2$ & $1500^2$ & $1500^2$\\
    & & & & & & & &\\
    \hline
\end{tabular}
\end{center}
\label{Bench_P1}
}
\end{table}

\begin{table}
\caption{Physical masses in GeV of the particles for the scenario of Table~\ref{Bench_P1}.}
\begin{center}
\begin{tabular}{|c|c|c|c|c|c|}
  \hline
  $\mnt{1}$ & $\mnt{2}$ & $\mnt{3}$ & $\mnt{4}$ & $\mch{1}$ & $\mch{2}$ \\
  \hline \hline
  $781$ & $882$ & $911$ & $1669$ & $782$ & $914$ \\
  \hline
\end{tabular}
\vskip 0.4cm
\begin{tabular}{|c|c|c|c|c|}
  \hline
  $m_{h^0}$ & $m_{H^0}$ & $m_{A^0}$ & $m_{H^+}$ \\
  \hline \hline
  $124$  & $5325$ & $5325$ & $5359$ \\
  \hline
\end{tabular}
\vskip 0.4cm
\begin{tabular}{|c|c|c|c|c|c|c|}
  \hline
  $\msg$ & $\msu{1}$ & $\msu{2}$ & $\msu{3}$ & $\msu{4}$ & $\msu{5}$ & $\msu{6}$ \\
  \hline \hline
  $4424$ & $868$ & $3011$ & $3331$ & $3877$ & $4402$ & $4402$ \\
  \hline
\end{tabular}
\vskip 0.4cm
\begin{tabular}{|c|c|c|c|c|c|}
  \hline
 $\msd{1}$ & $\msd{2}$ & $\msd{3}$ & $\msd{4}$ & $\msd{5}$ & $\msd{6}$ \\
  \hline \hline
  $1705$ & $2833$ & $3010$ & $3877$ & $4397$ & $4403$ \\
  \hline
\end{tabular}
\vskip 0.4cm
\begin{tabular}{|c|c|c|c|c|c|c|c|c|}
  \hline
  $m_{\sneut_1}$ &  $m_{\sneut_2}$ &  $m_{\sneut_3}$ &  $m_{\ti l_1}$ &  $m_{\ti l_2}$ 
                        &  $m_{\ti l_3}$ &  $m_{\ti l_4}$ &  $m_{\ti l_5}$ &  $m_{\ti l_6}$ \\
  \hline \hline
   $1509$ &  $1509$ &  $1528$ &  $1489$ &  $1489$ &  $1509$ &  $1512$ &  $1512$ &  $1545$ \\
  \hline
\end{tabular}
\end{center}
\label{physmasses}
\end{table}
%

%
%

\section{Conclusion}
%
We have studied the decays $h^0(125)\to c \bar c, b \bar b, 
b \bar s, \gamma \gamma, g g$ in the MSSM with general QFV. 
{\it For the first time}, we have performed the systematic MSSM parameter 
scan respecting all of the relevant theoretical and experimental 
constraints. 
{\it In strong contrast to} the usual studies in the MSSM with quark flavor 
conservation, we have found that the deviations of these MSSM decay widths from 
the SM values can be quite sizable. 
Future lepton colliders such as ILC, CLIC, CEPC, FCC-ee and MuC can 
observe such sizable deviations from the SM at high signal significance 
{\it even after} the failure of SUSY particle discovery at the HL-LHC.
In case the deviation pattern shown here is really observed at the lepton 
colliders, then it would strongly suggest the discovery of QFV SUSY (the 
MSSM with general QFV).

\end{document}